\documentclass[prl,reprint]{revtex4-1}

\usepackage{graphicx}
\usepackage{amsmath}
\usepackage{amssymb}
\usepackage{multirow}
\usepackage[margin=0.5in]{geometry}

\newcommand{\sub}[1]{\ensuremath{_{\textrm{#1}}}} 
\newcommand{\super}[1]{\ensuremath{^{\textrm{#1}}}} 
\newcommand{\sci}[2]{\ensuremath{#1 \times 10^{#2}}}

\renewcommand{\section}[1]{\emph{#1} - }

\begin{document}

\title{Spicing up continuum solvation models with SaLSA:\\the spherically-averaged liquid susceptibility \emph{ansatz}}

\author{Ravishankar Sundararaman, Kathleen A. Schwarz, Kendra Letchworth-Weaver, and T. A. Arias}
\affiliation{Cornell University Department of Physics, Ithaca, NY 14853, USA}
\date{\today}


\begin{abstract}
Continuum solvation models enable electronic structure calculations of systems
in liquid environments, but because of the large number of empirical parameters,
they are limited to the class of systems in their fit set (typically organic molecules).
Here, we derive a solvation model with no empirical parameters for the dielectric response
by taking the linear response limit of a classical density functional for molecular liquids.
This model directly incorporates the nonlocal dielectric response of the liquid
using an angular momentum expansion, and with a single fit parameter for
dispersion contributions it predicts solvation energies of neutral molecules with an
RMS error of 1.3 kcal/mol in water and 0.8 kcal/mol in chloroform and carbon tetrachloride.
We show that this model is more accurate for strongly polar and charged systems
than previous solvation models because of the parameter-free electric response,
and demonstrate its suitability for \emph{ab initio} solvation,
including self-consistent solvation in quantum Monte Carlo calculations.
\end{abstract}

\maketitle

Electronic density functional theory \cite{HK-DFT,KS-DFT} enables first-principles
prediction of material properties at the atomic scale including structures and
reaction mechanisms. Liquids play a vital role in many systems of
technological and scientific interest, but the need for thermodynamic
phase-space sampling complicates direct first-principles calculations.
Further, absence of dispersion interactions and neglect of quantum-mechanical
effects in the motion of protons limit the accuracy of \emph{ab initio}
molecular dynamics for solvents such as water \cite{Water-CPMD,Water-PIMD}.

Continuum solvation models replace the effect of the solvent
by the response of an empirically-determined dielectric cavity.
Traditional solvation models \cite{PCM94,PCM97,PCM-Review,PCM-SM1,PCM-SM8,PCM-SMD}
employ a large number of atom-dependent parameters, are highly accurate
in the class of systems to which they are fit - typically organic molecules in solution,
and have been tremendously successful in the evaluation of reaction mechanisms
and design of molecular catalysts. Unfortunately, the large number of parameters
precludes the extrapolation of these models to systems outside their fit set,
such as metallic or ionic surfaces in solution. Recent solvation models
that employ an electron-density based parametrization \cite{PCM-SCCS,PCM-Kendra}
require only two or three parameters and extrapolate more reliably,
but still encounter difficulties for charged and highly polar systems
\cite{PCM-SCCS-charged,NonlinearPCM}.

The need for empirical parameters in continuum solvation arises primarily
because of the drastic simplification of the nonlocal and nonlinear response
of the real liquid with that of a continuum dielectric cavity.
Recently, we correlated the dielectric cavity sizes for different solvents
with the extent of nonlocality of the solvent response to enable a
unified electron-density parametrization for multiple solvents \cite{CavityWDA},
but the electron density threshold $n_c$ that determines the cavity size
still required a fit to solvation energies of organic molecules.
Joint density functional theory (JDFT) \cite{JDFT} combines a classical density
functional description of the solvent with an electronic density functional
description of the solute, naturally captures the nonlocal response
of the fluid, and does not involve cavities that are fit to solvation energies.

Here, we derive a nonlocal continuum solvation model from the linear response
limit of JDFT. Because there are no fit parameters for the electrostatic response,
this theory is therefore suitable for the study of charged and strongly polar systems.
This derivation begins with a simple \emph{ansatz}: the distribution of solvent
molecules starts out isotropic and uniform outside a region excluded by the solute;
electrostatic interactions between the solute and solvent then perturb
this distribution to first order. We first describe a method to estimate
that initial distribution from the overlap of solute and solvent electron
densities, and then show how to calculate the nonlocal linear response
of the fluid using an angular momentum expansion. The addition of
nonlocal cavity formation free energy and dispersion functionals
derived from classical density functional theory \cite{CavityWDA}
results in an accurate description of solvation free energies
of neutral organic molecules as well as highly polar and ionic systems.
Finally, we show that the nonlocal dependence of this model on the
solute electron density enables self-consistent solvation in quantum
Monte-Carlo calculations, which was previously impractical due to
statistical noise in the local electron density \cite{Katie-QMC}.

\section{Iso-density-product cavity determination}
A common ingredient in continuum solvation models is the formation of a cavity
that excludes the solvent from a region of space occupied by the solute.
Atom-based parametrizations typically exclude the centers of solvent molecules
from a union of spheres centered on each solute atom with radius equal to the sum of
the atomic and solvent van-der-Waals (vdW) radii, and then construct a
dielectric cavity that is smaller by an empirical solvent-dependent radius \cite{PCM-Review}.
In contrast, the density-based approaches adopt a smoothly-varying dielectric constant
which is a function of the local solute electron density that switches from the vacuum
to bulk solvent value at a solvent-dependent critical electron density $n_c$ \cite{PCM-SCCS,PCM-Kendra}.

\begin{figure}
\centering{\includegraphics[width=0.8\columnwidth]{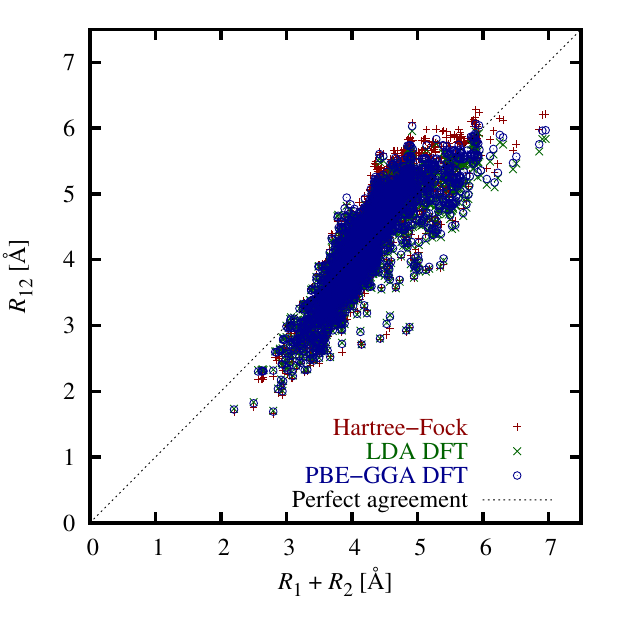}}
\caption{Atom separation, $R_{12}$, at which electron density overlap equals
$\bar{n}_c = \sci{1.42}{-3}\ a_0^{-3}$, compared to sum of van der Waals (vdW) radii,
$R_1 + R_2$, for all pairs of atoms with vdW radii tabulated in \cite{CRC-Handbook,RvdwAtoms}.
Differences between LDA, GGA and Hartree-Fock densities affect the agreement negligibly.
\label{fig:Rcompare}}
\end{figure}

Here, since we explicitly account for the nonlocal response,
we require only the distribution of the solvent molecule centers.
Because the distance of these centers from the solute atoms corresponds
to the distance of nearest approach of two non-bonded systems, 
vdW radii, which are defined in terms of non-bonded
approach distances \cite{RvdwAtoms}, provide a reasonable guess.
However, directly using the vdW radii does not account for changes
in the electronic configuration between the isolated atom and
molecules or solids. A description based on the electronic density
would instead naturally capture this dependence.

The interaction of non-bonded systems is dominated by Pauli repulsion
at short distances, which depends on the overlap of the electron
densities of the two systems.  Indeed, Figure~\ref{fig:Rcompare}
shows that the atom separation, $R_{12}$, at which the electron density overlap
$\bar{n}(R_{12}) = \int d\vec{r} n_1(\vec{r}) n_2(\vec{r})$ crosses a threshold
value of $\bar{n}_c = \sci{1.42}{-3}\ a_0^{-3}$ correlates well with
the sum of vdW radii $R_1 + R_2$ \cite{CRC-Handbook} of the two atoms.
Here, $n_1(\vec{r})$ and $n_2(\vec{r})$ are spherical electron densities
of the two atoms (calculated using OPIUM \cite{opium}), centered $R_{12}$ apart,
and we obtain $\bar{n}_c$ by minimizing $\sum_{ij} (R_i + R_j - R_{ij})^2$.
This result is insensitive to the choice of exchange-correlation approximation
used to calculate the densities: at the optimized $\bar{n}_c$, the RMS
relative error in $R_{12}$ compared to $R_1+R_2$ is $8.1\%$ for the
local-density approximation \cite{LDA-PZ}, $8.3\%$ for the PBE
generalized-gradient approximation \cite{PBE} and $9.9\%$ for Hartree-Fock theory.

The above analysis provides a universal threshold on the density product that can estimate
the approach distance of non-bonded systems. We utilize this capability to determine the
distribution of solvent molecule centers around a solute with electron density $n(\vec{r})$.
Our \emph{ansatz} requires a spatially-varying but isotropic initial distribution of molecules.
Therefore we compute overlaps with the spherically-averaged electron density $n\sub{lq}^0(r) = 
\int \frac{d\hat{n}}{4\pi} n\sub{lq}(r\hat{n})$, where $n\sub{lq}(\vec{r})$
is the electron density of a single solvent molecule and $\hat{n}$ is a unit vector.
Following \cite{NonlinearPCM}, we describe the spatial variation of the solvent distribution
by the smooth `cavity shape' functional
\begin{equation}
s(\vec{r}) = \frac{1}{2}\operatorname{erfc} \ln \frac{n\sub{lq}^0(r) \ast n(\vec{r})}{\bar{n}_c} \label{eqn:shapeSaLSA}
\end{equation}
which smoothly transitions from vacuum ($s=0$) to bulk fluid ($s=1$) as the overlap
of the solute and solvent electron densities (readily calculated as a convolution)
crosses the universal overlap threshold $\bar{n}_c$.

\section{Nonlocal electric response}
We begin with the in-principle exact joint density-functional description \cite{JDFT}
of the free energy of a solvated electronic system
\begin{equation}
A\sub{JDFT}[n,\{N_\alpha\}] = A\sub{HK}[n] + \Phi\sub{lq}[\{N_\alpha\}] + \Delta A[n,\{N_\alpha\}], \label{eqn:JDFT}
\end{equation}
where $A\sub{HK}$ is the Hohenberg-Kohn functional \cite{HK-DFT} of the solute electronic density $n(\vec{r})$,
$\Phi\sub{lq}$ is a free energy functional for the solvent in terms of nuclear site densities $\{N_\alpha(\vec{r})\}$,
and $\Delta A$ captures the free energy of interaction between the solute and solvent. (See \cite{JDFT} for details about the theoretical framework.)

We adopt the Kohn-Sham prescription \cite{KS-DFT} with an approximate exchange-correlation functional for $A\sub{HK}[n]$
and the polarizable `scalar-EOS' free energy functional approximation \cite{PolarizableCDFT} for $\Phi\sub{lq}[\{N_\alpha\}]$.
We separate $\Delta A$ in (\ref{eqn:JDFT}) as the mean-field electrostatic interaction and a remainder
that is dominated by electronic repulsion and dispersion interactions.
We then assume that the remainder is responsible for determining the
initial isotropic distribution $N_0(\vec{r}) = N\sub{bulk} s(\vec{r})$
of solvent molecules, where $N\sub{bulk}$ is the bulk number density
of solvent molecules and $s(\vec{r})$ is given by (\ref{eqn:shapeSaLSA}).
Substituting the free energy functional from \cite{PolarizableCDFT},
we can then write (\ref{eqn:JDFT}) as
\begin{widetext}
\begin{multline}
A\sub{JDFT} = A\sub{HK}[n] + \Phi_0[N_0]
	+ T \int d\vec{r} \int \frac{d\omega}{8\pi^2} p_\omega(\vec{r}) \left[ \ln \frac{p_\omega(\vec{r})}{N_0(\vec{r})} - 1 \right]
	+ \frac{C\sub{rot}^{-1}-1}{N\sub{bulk}p\sub{mol}^2/3T} \int d\vec{r} \left( \int \frac{d\omega}{8\pi^2} p_\omega(\vec{r}) \omega\circ\vec{p}\sub{mol} \right)^2 \\
	+ \sum_\alpha \int d\vec{r} \frac{N_\alpha(\vec{r}) \mathcal{P}_\alpha(\vec{r})^2}{2C\sub{pol}\chi_\alpha}
	+ \int d\vec{r} \int d\vec{r}' 
		\left( \rho\sub{el}(\vec{r}) + \frac{\rho\sub{lq}(\vec{r}) - \rho\sub{lq}^0(\vec{r})}{2} \right)
			\frac{1}{|\vec{r}-\vec{r}'|} (\rho\sub{lq}(\vec{r}') - \rho\sub{lq}^0(\vec{r}')). \label{eqn:JDFTsep}
\end{multline}
\end{widetext}
Above, analogously to the Kohn-Sham approach, the liquid free energy functional
employs the state of the corresponding non-interacting system
specified by two sets of independent variables. 
The first, $p_\omega(\vec{r})$, is the orientation probability density of finding
a solvent molecule centered at $\vec{r}$ with orientation $\omega\in$SO(3).
The solvent site densities $N_\alpha(\vec{r})$ are dependent variables that are calculated from $p_\omega(\vec{r})$.
The second variable is the polarization density, $\mathcal{P}_\alpha(\vec{r})$ for each solvent site.

The second term in (\ref{eqn:JDFTsep}), $\Phi_0[N_0]$, collects all the free energy contributions
due to the initial isotropic distribution, so that all the subsequent terms
are zero when $p_\omega(\vec{r})=N_0(\vec{r})$ and $\mathcal{P}_\alpha(\vec{r})=0$.
The third term is the non-interacting rotational entropy at temperature $T$,
the fourth term is a weighted-density correlation functional for dipole rotations,
and the fifth term is the potential energy for molecular polarization with site susceptibilities $\chi_\alpha$.
$C\sub{rot}$ and $C\sub{pol}$ parametrize correlations in the rotations and polarization respectively,
and are constrained by the bulk static and optical dielectric constants.
The final term is the mean-field interaction between the solute charge density $\rho\sub{el}(\vec{r})$
and the induced charge density in the liquid $\rho\sub{lq}(\vec{r}) - \rho\sub{lq}^0(\vec{r})$
(where $\rho\sub{lq}^0$ is the charge in the initial isotropic distribution),
and the self energy of that induced charge. Here,
\begin{equation}
\rho\sub{lq}(\vec{r}) = \sum_\alpha \rho_\alpha(r) \ast N_\alpha(\vec{r}) - \nabla\cdot \sum_\alpha w_\alpha(r) \ast N_\alpha \vec{\mathcal{P}}_\alpha, \label{eqn:SaLSA_rhoLq}
\end{equation}
where $\rho_\alpha(r)$ and $w_\alpha(r)$ respectively specify decomposition of the solvent molecule's
charge density and nonlocal susceptibility into spherical contributions at the solvent sites.
Bulk experimental properties of the liquid and \emph{ab initio} calculations of a single solvent molecule
constrain all involved parameters. See \cite{PolarizableCDFT} for details; the terms above are identical,
except for the inclusion of solute-solvent interactions in the final electrostatic term
and for the separation of the initial isotropic contributions into $\Phi_0[N_0]$.

Next, we treat the orientation-dependent pieces perturbatively by expanding
$p_\omega(\vec{r}) = N_0(\vec{r}) (1 + \sum_{lmm'} x^l_{mm'}(\vec{r}) D^l_{mm'}(\omega))$,
where $D^l_{mm'}(\omega)$ are the Wigner $D$-matrices (irreducible representations of SO(3)) \cite{GroupTheory-Wigner}.
We then expand the free energy (\ref{eqn:JDFTsep}) to quadratic order in the independent variables $x^l_{mm'}(\vec{r})$ (rotation)
and $\vec{\mathcal{P}}_\alpha(\vec{r})$ (polarization), formally solve the corresponding linear Euler-Lagrange equations
and substitute those solutions back into the quadratic form. After some tedious but straightforward algebra involving
orthogonality of $D$-matrices, addition of spherical harmonics and their transformation under the $D$-matrices,
as well as Fourier transforms to simplify convolutions, we can show that the resulting free energy to second order is exactly
\begin{multline}
A\sub{SaLSA} = A\sub{HK}[n] + \Phi_0[N_0] + \\
	\frac{1}{2} \int d\vec{r} \rho\sub{el}(\vec{r}) \left[ (\hat{K}^{-1} - \hat{\chi})^{-1} - \hat{K} \right] \rho\sub{el}(\vec{r}).
\label{eqn:SaLSA}
\end{multline}
Here, $\hat{K}$ is the Coulomb operator and $\hat{\chi}$ is the nonlocal `spherically-averaged liquid susceptibility'
(SaLSA), expressed conveniently in reciprocal space as
\begin{multline}
\hat{\chi}(\vec{G},\vec{G}') \equiv
	-\sum_\alpha \tilde{N}_\alpha^0(\vec{G}-\vec{G}') C\sub{pol}\chi_\alpha \vec{G}\cdot\vec{G}' \tilde{w}_\alpha(G) \tilde{w}^\ast_\alpha(G') \\
	-\tilde{N}_0(\vec{G}-\vec{G}') \sum_{lm} \frac{C\sub{rot}^l}{T} \frac{P_l(\hat{G}\cdot\hat{G}')}{4\pi} \tilde{\rho}\sub{mol}^{lm}(G) \tilde{\rho}\sub{mol}^{lm\ast}(G'), \label{eqn:chiSaLSA}
\end{multline}
where $\tilde{f}(\vec{G})$ is the Fourier transform of $f(\vec{r})$ for any $f$.
The first term of (\ref{eqn:chiSaLSA}) captures the polarization response, where
$N_\alpha^0(\vec{r}) = N_0(\vec{r}) \ast \delta(r-R_\alpha)/(4\pi R_\alpha^2)$
is the site density at the initial configuration $p_\omega = N_0(\vec{r})$
of a solvent site at a distance $R_\alpha$ from the solvent molecule center.
The second term of (\ref{eqn:chiSaLSA}) captures the rotational response
of solvent molecules with charge distribution $\rho\sub{mol}(\vec{r})$,
decomposed in Fourier space as $\tilde{\rho}\sub{mol}(\vec{G})
= \sum_{lm} \tilde{\rho}\sub{mol}^{lm}(G) Y_{lm}(\hat{G})$.
The prefactor $C\sub{rot}^l = C\sub{rot}$, the dipole rotation correlation
factor \cite{PolarizableCDFT} for $l=1$, and it equals unity for all other $l$.

\begin{figure}
\includegraphics[width=\columnwidth]{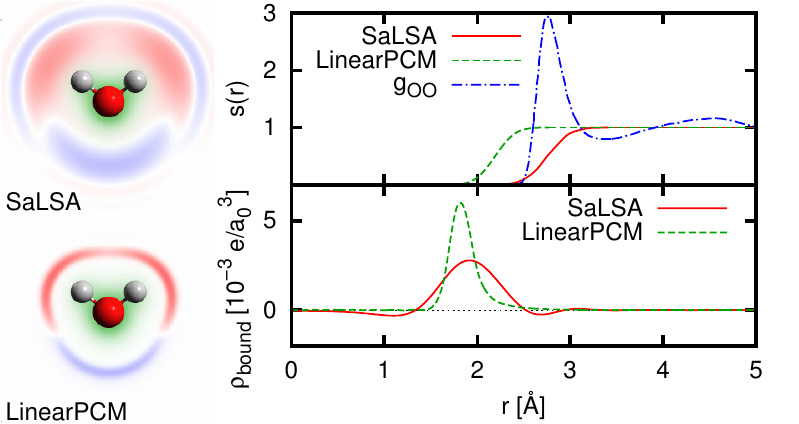}
\caption{Cavity shape function $s(\vec{r})$ and bound charge $\rho\sub{bound}(\vec{r})$
for a water molecule in water from SaLSA and the local LinearPCM model \cite{NonlinearPCM},
compared to the experimental oxygen-oxygen radial distribution function $g_{OO}(r)$ \cite{SoperEPSR}.
The left panels show the bound charge ($+$ red, $-$ blue) and electron density (green).
\label{fig:nbound}}
\end{figure}

For practical calculations, we rewrite the last term of (\ref{eqn:SaLSA}) as
$\int (\phi - \phi_0) \rho\sub{el}/2$, where $\phi_0 = \hat{K}\rho\sub{el}$ is the
electrostatic potential in vacuum and $\phi$ is the total (mean-field) electrostatic potential
which solves the modified Poisson-like equation $(\nabla^2 + 4\pi\hat{\chi}) \phi = -4\pi\rho\sub{el}$.
The $l=1$ rotational and polarization terms in $\hat{\chi} \phi$ have the structure
$\nabla\cdot w\ast N(\vec{r}) w\ast \nabla\phi$ which resembles the Poisson equation
for an inhomogeneous dielectric, except for the convolutions that introduce the nonlocality.
For neutral solvent molecules, the $l=0$ term captures the interaction of the solute
with a spherical charge distribution of zero net charge, and is zero except for small
contributions from non-zero but negligible overlap of the solute and solvent charges.
However, note that the SaLSA response easily generalizes to mixtures, and for ionic species
in the solution, the $l=0$ terms convert the Poisson-like equation to a Helmholtz-like
equation that naturally captures the Debye-screening effects of electrolytes as in \cite{Kendra-PCM}.
The $l>1$ terms resemble (nonlocal versions of) higher-order differential operators
and capture interactions with higher-order multipoles of the solvent molecule,
which decrease in magnitude with increasing $l$. We find that including terms
up to $l=3$ is sufficient to converge the solvation energies to 0.1~kcal/mol.

The nonlocality of the SaLSA response allows the fluid bound charge to appear
at a distance from the edge of the cavity. For example, for a water molecule
in liquid water, the SaLSA cavity transitions at about the first peak of the
radial distribution function $g_{OO}(r)$ of water (Figure~\ref{fig:nbound}),
whereas the bound charge is dominant at smaller distances. In contrast,
local solvation models require a smaller unphysical cavity to produce
bound charge at the appropriate distance to capture the experimental solvation energy.
This key difference from the local models allows a
non-empirical description of the electric response in SaLSA.

At this stage, the solvated free energy (\ref{eqn:SaLSA}) is fully specified
except for the free energy of the initial configuration $\Phi_0[N_0]$,
dominated by electronic repulsion, dispersion and the free energy of
forming a cavity in the liquid. We set $\Phi_0[N_0] = G\sub{cav}[s] + E\sub{disp}[N_0]$,
where $G\sub{cav}$ is a non-local weighted density approximation to the
cavity formation free energy and $E\sub{disp}$ empirically accounts
for dispersion and the remaining contributions, exactly as in \cite{CavityWDA}.
(See \cite{CavityWDA} for a full specification.) Briefly, $G\sub{cav}$ is completely
constrained by bulk properties of the solvent including density, surface tension
and vapor pressure, and reproduces the classical density functional and
molecular dynamics predictions for the cavity formation free energy from \cite{PolarizableCDFT}
with no fit parameters. $E\sub{disp}$ employs a semi-empirical pair potential
dispersion correction \cite{Dispersion-Grimme} which includes a scale parameter $s_6$,
which we fit to solvation energies below. Note, however, that unlike previous
continuum solvation models, the dominant electrostatic response includes
no parameters that are fit to solvation energies.

\section{Solvation energies}
We implement the SaLSA solvation model in the open source density-functional software JDFTx \cite{JDFTx},
and perform calculations using norm-conserving pseudopotentials \cite{opium} at 30~$E_h$ plane-wave cutoff
and the revTPSS meta-generalized-gradient exchange-correlation functional \cite{revTPSS}.
For three solvents, water, chloroform and carbon tetrachloride, we fit the sole parameter
$s_6$ to minimize the RMS error in the solvation energies of a small but representative set
of neutral organic molecules with a variety of functional groups and chain lengths
(same set for each solvent as in \cite{CavityWDA}). Table~\ref{tab:fitParamsSaLSA}
summarizes the optimum values of $s_6$ and the corresponding RMS error in solvation energies.
The RMS errors of SaLSA are only slightly higher than those of the local solvation model
from \cite{CavityWDA} that includes additional fit parameters for the electric response.

\begin{table}
\caption{Fit parameter and residual for the SaLSA nonlocal solvation model.
All other quantities for these solvents are obtained from bulk data
and \emph{ab initio} calculations and are listed in \cite{PolarizableCDFT}.
\label{tab:fitParamsSaLSA}}
\begin{center}\begin{tabular}{c|ccc}
\hline\hline
\multirow{2}{*}{Solvent} & \multirow{2}{*}{$s_6$} & \multicolumn{2}{c}{RMS error [kcal/mol (m$E_h$)]} \\
& & SaLSA & Local model \cite{CavityWDA} \\
\hline
H\sub{2}O    & 0.50 & 1.3 (2.0) & 1.1 (1.8) \\
CHCl\sub{3}  & 0.88 & 0.7 (1.1) & 0.6 (1.0) \\
CCl\sub{4}   & 1.06 & 0.8 (1.3) & 0.5 (0.8) \\
\hline\hline
\end{tabular}\end{center}
\end{table}

\begin{figure}
\includegraphics[width=\columnwidth]{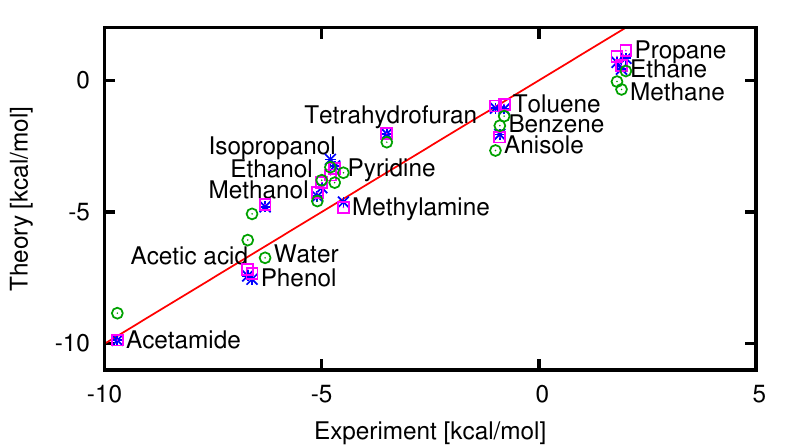}\\
\includegraphics[width=\columnwidth]{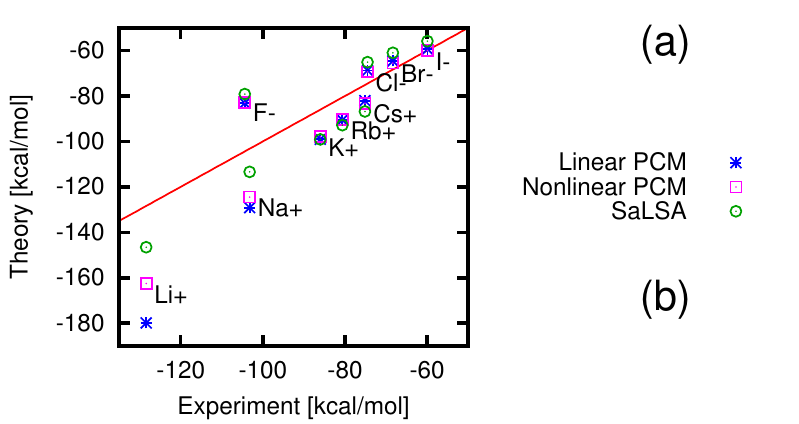}
\caption{Solvation energies in water predicted by the SaLSA nonlocal solvation model for 
(a) organic molecules and (b) ions compared to experiment \cite{solvation-exp-compiled-1,solvation-exp-compiled-2,solvation-exp-ions}
and the linear and nonlinear solvation models from \cite{NonlinearPCM}.
\label{fig:SolvationEnergies}}
\end{figure}

Figure~\ref{fig:SolvationEnergies} compares the aqueous solvation energy predictions
of SaLSA with those of the linear and nonlinear local-response models from \cite{NonlinearPCM}.
All three models perform comparably for the neutral molecule set (Figure~\ref{fig:SolvationEnergies}(a))
used for the parameter fit, but SaLSA is substantially more accurate for inorganic ions
(Figure~\ref{fig:SolvationEnergies}(b)). In particular, the local models severely over-predict
the solvation energies of small cations, and the nonlocal SaLSA model reduces the error
by a factor of three for Li+ and Na+. However, SaLSA does not correct the systematic
over-solvation of cations compared to anions, a known deficiency of
electron-density based solvation models \cite{PCM-SCCS-charged}.

Finally, we turn to solvation in diffusion quantum Monte-Carlo (DMC) calculations.
Conventional density-based solvation models \cite{PCM-SCCS,PCM-SCCS-charged,PCM-Kendra,NonlinearPCM}
are sensitive to the electron density in the low density regions of space
($n(\vec{r}) \sim 10^{-4}-10^{-3}$~$a_0^{-3}$).
This sensitivity imposes extremely stringent restrictions on the statistical noise in the
DMC electron density, rendering self-consistent solvated DMC calculations impractical.
A scheme correct to first order that combines a solvated density-functional calculation
with a DMC calculation in an external potential provides reasonable accuracy
without calculating DMC electron densities \cite{Katie-QMC}. However, full self-consistency
would be particularly important for systems where density-functional approximations fail drastically.

\begin{figure}
\includegraphics[width=\columnwidth]{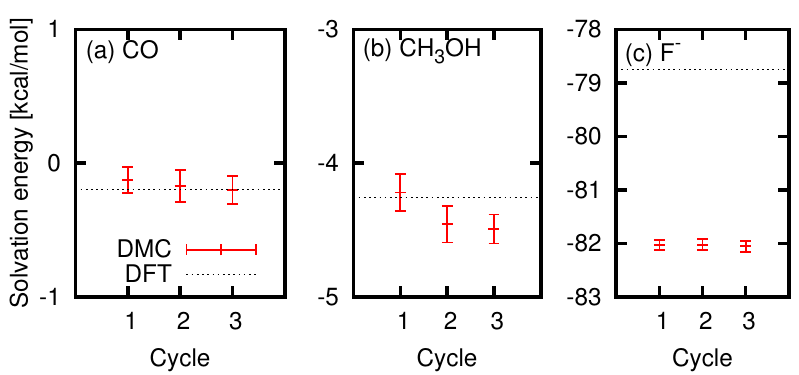}
\caption{Convergence in self-consistent diffusion Monte-Carlo solvation
energies for (a) carbon monoxide, (b) methanol and (c) fluoride anion in water.
\label{fig:qmcSol}}
\end{figure}

The nonlocality of the SaLSA model, particularly the dependence of the cavity shape
(\ref{eqn:shapeSaLSA}) on a convolution of the density rather than the local density,
significantly reduces its sensitivity to noise in the electron density, and enables
self-consistent DMC solvation in a straightforward manner.
We start with an initial guess for the density (from solvated DFT),
compute the potential on the electrons from the SaLSA fluid model, and
perform a DMC calculation (using the CASINO program \cite{CASINO})
in that external potential while collecting electron density (using the mixed estimator in \cite{CASINO}).
We then mix the DMC electron density with the previous guess, update the
fluid potential and repeat till the density becomes self-consistent.
The estimation of the solvated energy at each cycle proceeds exactly as in
\cite{Katie-QMC} with slight differences in the details of the DMC calculation:
we use Trail-Needs pseudopotentials \cite{TrailNeeds,TN2} 
with a DFT plane-wave cutoff of 70~$E_h$, and a DMC time step of 0.004~$E_h^{-1}$.

Figure~\ref{fig:qmcSol} shows that the solvated DMC energy converges
to within 0.1~kcal/mol in just three self-consistency cycles. The solvation energies of
methanol and carbon monoxide are essentially unchanged from the
density-functional results, whereas the solvation energy of the fluoride
anion, which suffers from strong self-interaction errors in DFT,
is corrected by 3~kcal/mol. This change, although in the right direction,
is small compared to the 20~kcal/mol error in the predicted SaLSA
solvation energy using DFT (Figure~\ref{fig:SolvationEnergies}(b)).
The error in the solvation of anions is therefore not predominantly
caused by the inaccuracy of electronic density-functional approximations for anions.

\section{Conclusions}
The linear-response limit of joint density-functional theory, combined with an
electron-density overlap based estimate of the initial solvent molecule distribution,
provides a nonlocal continuum solvation model with no empirical parameters for the electric response.
Consequently, this `SaLSA' model extrapolates more reliably from neutral organic molecules to ions
and is an excellent candidate for describing highly polar and charged surfaces in solution.
Further, the nonlocality of this model enables self-consistent solvation in diffusion Monte Carlo calculations.
This opens up the possibility of studying systems in solution for which standard density-functional
approximations fail, such as the adsorption of carbon monoxide on transition metal surfaces.
Additionally, SaLSA should facilitate the development of more accurate and perhaps more empirical
models that, for example, account for the charge asymmetry in the solvation of cations and anions.

This work was supported as a part of the Energy Materials Center at Cornell (EMC$^2$),
an Energy Frontier Research Center funded by the U.S. Department of Energy,
Office of Science, Office of Basic Energy Sciences under Award Number DE-SC0001086.


\begin{thebibliography}{34}%
\makeatletter
\providecommand \@ifxundefined [1]{%
 \@ifx{#1\undefined}
}%
\providecommand \@ifnum [1]{%
 \ifnum #1\expandafter \@firstoftwo
 \else \expandafter \@secondoftwo
 \fi
}%
\providecommand \@ifx [1]{%
 \ifx #1\expandafter \@firstoftwo
 \else \expandafter \@secondoftwo
 \fi
}%
\providecommand \natexlab [1]{#1}%
\providecommand \enquote  [1]{``#1''}%
\providecommand \bibnamefont  [1]{#1}%
\providecommand \bibfnamefont [1]{#1}%
\providecommand \citenamefont [1]{#1}%
\providecommand \href@noop [0]{\@secondoftwo}%
\providecommand \href [0]{\begingroup \@sanitize@url \@href}%
\providecommand \@href[1]{\@@startlink{#1}\@@href}%
\providecommand \@@href[1]{\endgroup#1\@@endlink}%
\providecommand \@sanitize@url [0]{\catcode `\\12\catcode `\$12\catcode
  `\&12\catcode `\#12\catcode `\^12\catcode `\_12\catcode `\%12\relax}%
\providecommand \@@startlink[1]{}%
\providecommand \@@endlink[0]{}%
\providecommand \url  [0]{\begingroup\@sanitize@url \@url }%
\providecommand \@url [1]{\endgroup\@href {#1}{\urlprefix }}%
\providecommand \urlprefix  [0]{URL }%
\providecommand \Eprint [0]{\href }%
\providecommand \doibase [0]{http://dx.doi.org/}%
\providecommand \selectlanguage [0]{\@gobble}%
\providecommand \bibinfo  [0]{\@secondoftwo}%
\providecommand \bibfield  [0]{\@secondoftwo}%
\providecommand \translation [1]{[#1]}%
\providecommand \BibitemOpen [0]{}%
\providecommand \bibitemStop [0]{}%
\providecommand \bibitemNoStop [0]{.\EOS\space}%
\providecommand \EOS [0]{\spacefactor3000\relax}%
\providecommand \BibitemShut  [1]{\csname bibitem#1\endcsname}%
\let\auto@bib@innerbib\@empty
\bibitem [{\citenamefont {Hohenberg}\ and\ \citenamefont
  {Kohn}(1964)}]{HK-DFT}%
  \BibitemOpen
  \bibfield  {author} {\bibinfo {author} {\bibfnamefont {P.}~\bibnamefont
  {Hohenberg}}\ and\ \bibinfo {author} {\bibfnamefont {W.}~\bibnamefont
  {Kohn}},\ }\href@noop {} {\bibfield  {journal} {\bibinfo  {journal} {Phys.
  Rev.}\ }\textbf {\bibinfo {volume} {136}},\ \bibinfo {pages} {B864} (\bibinfo
  {year} {1964})}\BibitemShut {NoStop}%
\bibitem [{\citenamefont {Kohn}\ and\ \citenamefont {Sham}(1965)}]{KS-DFT}%
  \BibitemOpen
  \bibfield  {author} {\bibinfo {author} {\bibfnamefont {W.}~\bibnamefont
  {Kohn}}\ and\ \bibinfo {author} {\bibfnamefont {L.}~\bibnamefont {Sham}},\
  }\href@noop {} {\bibfield  {journal} {\bibinfo  {journal} {Phys. Rev.}\
  }\textbf {\bibinfo {volume} {140}},\ \bibinfo {pages} {A1133} (\bibinfo
  {year} {1965})}\BibitemShut {NoStop}%
\bibitem [{\citenamefont {Grossman}\ \emph {et~al.}(2004)\citenamefont
  {Grossman}, \citenamefont {Schwegler}, \citenamefont {Draeger}, \citenamefont
  {Gygi},\ and\ \citenamefont {Galli}}]{Water-CPMD}%
  \BibitemOpen
  \bibfield  {author} {\bibinfo {author} {\bibfnamefont {J.~C.}\ \bibnamefont
  {Grossman}}, \bibinfo {author} {\bibfnamefont {E.}~\bibnamefont {Schwegler}},
  \bibinfo {author} {\bibfnamefont {E.~W.}\ \bibnamefont {Draeger}}, \bibinfo
  {author} {\bibfnamefont {F.}~\bibnamefont {Gygi}}, \ and\ \bibinfo {author}
  {\bibfnamefont {G.}~\bibnamefont {Galli}},\ }\href@noop {} {\bibfield
  {journal} {\bibinfo  {journal} {J. Chem. Phys.}\ }\textbf {\bibinfo {volume}
  {120}},\ \bibinfo {pages} {300} (\bibinfo {year} {2004})}\BibitemShut
  {NoStop}%
\bibitem [{\citenamefont {Morrone}\ and\ \citenamefont
  {Car}(2008)}]{Water-PIMD}%
  \BibitemOpen
  \bibfield  {author} {\bibinfo {author} {\bibfnamefont {J.~A.}\ \bibnamefont
  {Morrone}}\ and\ \bibinfo {author} {\bibfnamefont {R.}~\bibnamefont {Car}},\
  }\href@noop {} {\bibfield  {journal} {\bibinfo  {journal} {Phys. Rev. Lett.}\
  }\textbf {\bibinfo {volume} {101}},\ \bibinfo {pages} {017801} (\bibinfo
  {year} {2008})}\BibitemShut {NoStop}%
\bibitem [{\citenamefont {Fortunelli}\ and\ \citenamefont
  {Tomasi}(1994)}]{PCM94}%
  \BibitemOpen
  \bibfield  {author} {\bibinfo {author} {\bibfnamefont {A.}~\bibnamefont
  {Fortunelli}}\ and\ \bibinfo {author} {\bibfnamefont {J.}~\bibnamefont
  {Tomasi}},\ }\href@noop {} {\bibfield  {journal} {\bibinfo  {journal} {Chem.
  Phys. Lett.}\ }\textbf {\bibinfo {volume} {231}},\ \bibinfo {pages} {34}
  (\bibinfo {year} {1994})}\BibitemShut {NoStop}%
\bibitem [{\citenamefont {Barone}\ \emph {et~al.}(1997)\citenamefont {Barone},
  \citenamefont {Cossi},\ and\ \citenamefont {Tomasi}}]{PCM97}%
  \BibitemOpen
  \bibfield  {author} {\bibinfo {author} {\bibfnamefont {V.}~\bibnamefont
  {Barone}}, \bibinfo {author} {\bibfnamefont {M.}~\bibnamefont {Cossi}}, \
  and\ \bibinfo {author} {\bibfnamefont {J.}~\bibnamefont {Tomasi}},\
  }\href@noop {} {\bibfield  {journal} {\bibinfo  {journal} {J. Chem. Phys.}\
  }\textbf {\bibinfo {volume} {107}},\ \bibinfo {pages} {3210} (\bibinfo {year}
  {1997})}\BibitemShut {NoStop}%
\bibitem [{\citenamefont {Tomasi}\ \emph {et~al.}(2005)\citenamefont {Tomasi},
  \citenamefont {Mennucci},\ and\ \citenamefont {Cammi}}]{PCM-Review}%
  \BibitemOpen
  \bibfield  {author} {\bibinfo {author} {\bibfnamefont {J.}~\bibnamefont
  {Tomasi}}, \bibinfo {author} {\bibfnamefont {B.}~\bibnamefont {Mennucci}}, \
  and\ \bibinfo {author} {\bibfnamefont {R.}~\bibnamefont {Cammi}},\
  }\href@noop {} {\bibfield  {journal} {\bibinfo  {journal} {Chem. Rev.}\
  }\textbf {\bibinfo {volume} {105}},\ \bibinfo {pages} {2999} (\bibinfo {year}
  {2005})}\BibitemShut {NoStop}%
\bibitem [{\citenamefont {Cramer}\ and\ \citenamefont
  {Truhlar}(1991)}]{PCM-SM1}%
  \BibitemOpen
  \bibfield  {author} {\bibinfo {author} {\bibfnamefont {C.~J.}\ \bibnamefont
  {Cramer}}\ and\ \bibinfo {author} {\bibfnamefont {D.~G.}\ \bibnamefont
  {Truhlar}},\ }\href@noop {} {\bibfield  {journal} {\bibinfo  {journal} {J.Am.
  Chem. Soc.}\ }\textbf {\bibinfo {volume} {113}},\ \bibinfo {pages} {8305}
  (\bibinfo {year} {1991})}\BibitemShut {NoStop}%
\bibitem [{\citenamefont {Marenich}\ \emph {et~al.}(2007)\citenamefont
  {Marenich}, \citenamefont {Olson}, \citenamefont {Kelly}, \citenamefont
  {Cramer},\ and\ \citenamefont {Truhlar}}]{PCM-SM8}%
  \BibitemOpen
  \bibfield  {author} {\bibinfo {author} {\bibfnamefont {A.~V.}\ \bibnamefont
  {Marenich}}, \bibinfo {author} {\bibfnamefont {R.~M.}\ \bibnamefont {Olson}},
  \bibinfo {author} {\bibfnamefont {C.~P.}\ \bibnamefont {Kelly}}, \bibinfo
  {author} {\bibfnamefont {C.~J.}\ \bibnamefont {Cramer}}, \ and\ \bibinfo
  {author} {\bibfnamefont {D.~G.}\ \bibnamefont {Truhlar}},\ }\href@noop {}
  {\bibfield  {journal} {\bibinfo  {journal} {J. Chem. Theory Comput.}\ ,\
  \bibinfo {pages} {2011}} (\bibinfo {year} {2007})}\BibitemShut {NoStop}%
\bibitem [{\citenamefont {Marenich}\ \emph {et~al.}(2009)\citenamefont
  {Marenich}, \citenamefont {Cramer},\ and\ \citenamefont {Truhlar}}]{PCM-SMD}%
  \BibitemOpen
  \bibfield  {author} {\bibinfo {author} {\bibfnamefont {A.~V.}\ \bibnamefont
  {Marenich}}, \bibinfo {author} {\bibfnamefont {C.~J.}\ \bibnamefont
  {Cramer}}, \ and\ \bibinfo {author} {\bibfnamefont {D.~G.}\ \bibnamefont
  {Truhlar}},\ }\href@noop {} {\bibfield  {journal} {\bibinfo  {journal} {J.
  Phys. Chem. B}\ }\textbf {\bibinfo {volume} {113}},\ \bibinfo {pages} {6378}
  (\bibinfo {year} {2009})}\BibitemShut {NoStop}%
\bibitem [{\citenamefont {Andreussi}\ \emph {et~al.}(2012)\citenamefont
  {Andreussi}, \citenamefont {Dabo},\ and\ \citenamefont {Marzari}}]{PCM-SCCS}%
  \BibitemOpen
  \bibfield  {author} {\bibinfo {author} {\bibfnamefont {O.}~\bibnamefont
  {Andreussi}}, \bibinfo {author} {\bibfnamefont {I.}~\bibnamefont {Dabo}}, \
  and\ \bibinfo {author} {\bibfnamefont {N.}~\bibnamefont {Marzari}},\
  }\href@noop {} {\bibfield  {journal} {\bibinfo  {journal} {J. Chem. Phys}\
  }\textbf {\bibinfo {volume} {136}},\ \bibinfo {pages} {064102} (\bibinfo
  {year} {2012})}\BibitemShut {NoStop}%
\bibitem [{\citenamefont {Letchworth-Weaver}\ and\ \citenamefont
  {Arias}(2012)}]{PCM-Kendra}%
  \BibitemOpen
  \bibfield  {author} {\bibinfo {author} {\bibfnamefont {K.}~\bibnamefont
  {Letchworth-Weaver}}\ and\ \bibinfo {author} {\bibfnamefont {T.~A.}\
  \bibnamefont {Arias}},\ }\href@noop {} {\bibfield  {journal} {\bibinfo
  {journal} {Phys. Rev. B}\ }\textbf {\bibinfo {volume} {86}},\ \bibinfo
  {pages} {075140} (\bibinfo {year} {2012})}\BibitemShut {NoStop}%
\bibitem [{\citenamefont {Dupont}\ \emph {et~al.}(2013)\citenamefont {Dupont},
  \citenamefont {Andreussi},\ and\ \citenamefont {Marzari}}]{PCM-SCCS-charged}%
  \BibitemOpen
  \bibfield  {author} {\bibinfo {author} {\bibfnamefont {C.}~\bibnamefont
  {Dupont}}, \bibinfo {author} {\bibfnamefont {O.}~\bibnamefont {Andreussi}}, \
  and\ \bibinfo {author} {\bibfnamefont {N.}~\bibnamefont {Marzari}},\
  }\href@noop {} {\bibfield  {journal} {\bibinfo  {journal} {J. Chem. Phys}\
  }\textbf {\bibinfo {volume} {139}},\ \bibinfo {pages} {214110} (\bibinfo
  {year} {2013})}\BibitemShut {NoStop}%
\bibitem [{\citenamefont {Gunceler}\ \emph {et~al.}(2013)\citenamefont
  {Gunceler}, \citenamefont {Letchworth-Weaver}, \citenamefont {Sundararaman},
  \citenamefont {Schwarz},\ and\ \citenamefont {Arias}}]{NonlinearPCM}%
  \BibitemOpen
  \bibfield  {author} {\bibinfo {author} {\bibfnamefont {D.}~\bibnamefont
  {Gunceler}}, \bibinfo {author} {\bibfnamefont {K.}~\bibnamefont
  {Letchworth-Weaver}}, \bibinfo {author} {\bibfnamefont {R.}~\bibnamefont
  {Sundararaman}}, \bibinfo {author} {\bibfnamefont {K.}~\bibnamefont
  {Schwarz}}, \ and\ \bibinfo {author} {\bibfnamefont {T.}~\bibnamefont
  {Arias}},\ }\href@noop {} {\bibfield  {journal} {\bibinfo  {journal}
  {Modelling Simul. Mater. Sci. Eng.}\ }\textbf {\bibinfo {volume} {21}},\
  \bibinfo {pages} {074005} (\bibinfo {year} {2013})}\BibitemShut {NoStop}%
\bibitem [{\citenamefont {Sundararaman}\ \emph {et~al.}()\citenamefont
  {Sundararaman}, \citenamefont {Gunceler},\ and\ \citenamefont
  {Arias}}]{CavityWDA}%
  \BibitemOpen
  \bibfield  {author} {\bibinfo {author} {\bibfnamefont {R.}~\bibnamefont
  {Sundararaman}}, \bibinfo {author} {\bibfnamefont {D.}~\bibnamefont
  {Gunceler}}, \ and\ \bibinfo {author} {\bibfnamefont {T.~A.}\ \bibnamefont
  {Arias}},\ }\href@noop {} {\enquote {\bibinfo {title} {Weighted-density
  functionals for cavity formation and dispersion energies in continuum
  solvation models},}\ }\bibinfo {note} {Preprint arXiv:1407.4011}\BibitemShut
  {NoStop}%
\bibitem [{\citenamefont {Petrosyan}\ \emph {et~al.}(2007)\citenamefont
  {Petrosyan}, \citenamefont {Briere}, \citenamefont {Roundy},\ and\
  \citenamefont {Arias}}]{JDFT}%
  \BibitemOpen
  \bibfield  {author} {\bibinfo {author} {\bibfnamefont {S.~A.}\ \bibnamefont
  {Petrosyan}}, \bibinfo {author} {\bibfnamefont {J.-F.}\ \bibnamefont
  {Briere}}, \bibinfo {author} {\bibfnamefont {D.}~\bibnamefont {Roundy}}, \
  and\ \bibinfo {author} {\bibfnamefont {T.~A.}\ \bibnamefont {Arias}},\
  }\href@noop {} {\bibfield  {journal} {\bibinfo  {journal} {Phys. Rev. B}\
  }\textbf {\bibinfo {volume} {75}},\ \bibinfo {pages} {205105} (\bibinfo
  {year} {2007})}\BibitemShut {NoStop}%
\bibitem [{\citenamefont {Schwarz}\ \emph {et~al.}(2012)\citenamefont
  {Schwarz}, \citenamefont {Sundararaman}, \citenamefont {Letchworth-Weaver},
  \citenamefont {Arias},\ and\ \citenamefont {Hennig}}]{Katie-QMC}%
  \BibitemOpen
  \bibfield  {author} {\bibinfo {author} {\bibfnamefont {K.~A.}\ \bibnamefont
  {Schwarz}}, \bibinfo {author} {\bibfnamefont {R.}~\bibnamefont
  {Sundararaman}}, \bibinfo {author} {\bibfnamefont {K.}~\bibnamefont
  {Letchworth-Weaver}}, \bibinfo {author} {\bibfnamefont {T.~A.}\ \bibnamefont
  {Arias}}, \ and\ \bibinfo {author} {\bibfnamefont {R.~G.}\ \bibnamefont
  {Hennig}},\ }\href@noop {} {\bibfield  {journal} {\bibinfo  {journal} {Phys.
  Rev. B}\ }\textbf {\bibinfo {volume} {85}},\ \bibinfo {pages} {201102(R)}
  (\bibinfo {year} {2012})}\BibitemShut {NoStop}%
\bibitem [{\citenamefont {Haynes}(2012)}]{CRC-Handbook}%
  \BibitemOpen
  \bibinfo {editor} {\bibfnamefont {W.~M.}\ \bibnamefont {Haynes}},\ ed.,\
  \href@noop {} {\emph {\bibinfo {title} {CRC Handbook of Physics and Chemistry
  93\super{rd} ed}}}\ (\bibinfo  {publisher} {Taylor and Francis},\ \bibinfo
  {year} {2012})\BibitemShut {NoStop}%
\bibitem [{\citenamefont {Mantina}\ \emph {et~al.}(2009)\citenamefont
  {Mantina}, \citenamefont {Chamberlin}, \citenamefont {Valero}, \citenamefont
  {Cramer},\ and\ \citenamefont {Truhlar}}]{RvdwAtoms}%
  \BibitemOpen
  \bibfield  {author} {\bibinfo {author} {\bibfnamefont {M.}~\bibnamefont
  {Mantina}}, \bibinfo {author} {\bibfnamefont {A.~C.}\ \bibnamefont
  {Chamberlin}}, \bibinfo {author} {\bibfnamefont {R.}~\bibnamefont {Valero}},
  \bibinfo {author} {\bibfnamefont {C.~J.}\ \bibnamefont {Cramer}}, \ and\
  \bibinfo {author} {\bibfnamefont {D.~G.}\ \bibnamefont {Truhlar}},\
  }\href@noop {} {\bibfield  {journal} {\bibinfo  {journal} {J. Phys. Chem. A}\
  }\textbf {\bibinfo {volume} {113}},\ \bibinfo {pages} {5806} (\bibinfo {year}
  {2009})}\BibitemShut {NoStop}%
\bibitem [{\citenamefont {{OPIUM}}()}]{opium}%
  \BibitemOpen
  \bibfield  {author} {\bibinfo {author} {\bibnamefont {{OPIUM}}},\ }\href@noop
  {} {\emph {\bibinfo {title} {Pseudopotential generation project}}},\ \bibinfo
  {note} {\url{http://opium.sf.net}}\BibitemShut {NoStop}%
\bibitem [{\citenamefont {Perdew}\ and\ \citenamefont {Zunger}(1981)}]{LDA-PZ}%
  \BibitemOpen
  \bibfield  {author} {\bibinfo {author} {\bibnamefont {Perdew}}\ and\ \bibinfo
  {author} {\bibnamefont {Zunger}},\ }\href@noop {} {\bibfield  {journal}
  {\bibinfo  {journal} {Phys. Rev. B}\ }\textbf {\bibinfo {volume} {23}},\
  \bibinfo {pages} {5048} (\bibinfo {year} {1981})}\BibitemShut {NoStop}%
\bibitem [{\citenamefont {Perdew}\ \emph {et~al.}(1996)\citenamefont {Perdew},
  \citenamefont {Burke},\ and\ \citenamefont {Ernzerhof}}]{PBE}%
  \BibitemOpen
  \bibfield  {author} {\bibinfo {author} {\bibfnamefont {J.~P.}\ \bibnamefont
  {Perdew}}, \bibinfo {author} {\bibfnamefont {K.}~\bibnamefont {Burke}}, \
  and\ \bibinfo {author} {\bibfnamefont {M.}~\bibnamefont {Ernzerhof}},\
  }\href@noop {} {\bibfield  {journal} {\bibinfo  {journal} {Phys. Rev. Lett.}\
  }\textbf {\bibinfo {volume} {77}},\ \bibinfo {pages} {3865} (\bibinfo {year}
  {1996})}\BibitemShut {NoStop}%
\bibitem [{\citenamefont {Sundararaman}\ \emph {et~al.}(2014)\citenamefont
  {Sundararaman}, \citenamefont {Letchworth-Weaver},\ and\ \citenamefont
  {Arias}}]{PolarizableCDFT}%
  \BibitemOpen
  \bibfield  {author} {\bibinfo {author} {\bibfnamefont {R.}~\bibnamefont
  {Sundararaman}}, \bibinfo {author} {\bibfnamefont {K.}~\bibnamefont
  {Letchworth-Weaver}}, \ and\ \bibinfo {author} {\bibfnamefont {T.~A.}\
  \bibnamefont {Arias}},\ }\href@noop {} {\bibfield  {journal} {\bibinfo
  {journal} {J . Chem. Phys.}\ }\textbf {\bibinfo {volume} {140}},\ \bibinfo
  {pages} {144504} (\bibinfo {year} {2014})}\BibitemShut {NoStop}%
\bibitem [{\citenamefont {Wigner}(1959)}]{GroupTheory-Wigner}%
  \BibitemOpen
  \bibfield  {author} {\bibinfo {author} {\bibfnamefont {E.~P.}\ \bibnamefont
  {Wigner}},\ }\href@noop {} {\emph {\bibinfo {title} {Group theory and its
  application to the quantum mechanics of atomic spectra}}}\ (\bibinfo
  {publisher} {Academic Press, New York},\ \bibinfo {year} {1959})\BibitemShut
  {NoStop}%
\bibitem [{\citenamefont {Soper}(2000)}]{SoperEPSR}%
  \BibitemOpen
  \bibfield  {author} {\bibinfo {author} {\bibfnamefont {A.~K.}\ \bibnamefont
  {Soper}},\ }\href@noop {} {\bibfield  {journal} {\bibinfo  {journal} {Chem.
  Phys.}\ }\textbf {\bibinfo {volume} {258}},\ \bibinfo {pages} {121} (\bibinfo
  {year} {2000})}\BibitemShut {NoStop}%
\bibitem [{\citenamefont {Grimme}(2006)}]{Dispersion-Grimme}%
  \BibitemOpen
  \bibfield  {author} {\bibinfo {author} {\bibfnamefont {S.}~\bibnamefont
  {Grimme}},\ }\href@noop {} {\bibfield  {journal} {\bibinfo  {journal} {J.
  Comput. Chem}\ }\textbf {\bibinfo {volume} {27}},\ \bibinfo {pages} {1787}
  (\bibinfo {year} {2006})}\BibitemShut {NoStop}%
\bibitem [{\citenamefont {Sundararaman}\ \emph {et~al.}(2012)\citenamefont
  {Sundararaman}, \citenamefont {D.~Gunceler},\ and\ \citenamefont
  {Arias}}]{JDFTx}%
  \BibitemOpen
  \bibfield  {author} {\bibinfo {author} {\bibfnamefont {R.}~\bibnamefont
  {Sundararaman}}, \bibinfo {author} {\bibfnamefont {K.~L.-W.}\ \bibnamefont
  {D.~Gunceler}}, \ and\ \bibinfo {author} {\bibfnamefont {T.~A.}\ \bibnamefont
  {Arias}},\ }\href@noop {} {\enquote {\bibinfo {title} {{JDFTx}},}\ }\bibinfo
  {howpublished} {\url{http://jdftx.sourceforge.net}} (\bibinfo {year}
  {2012})\BibitemShut {NoStop}%
\bibitem [{\citenamefont {Perdew}\ \emph {et~al.}(2009)\citenamefont {Perdew},
  \citenamefont {Ruzsinszky}, \citenamefont {Csonka}, \citenamefont
  {Constantin},\ and\ \citenamefont {Sun}}]{revTPSS}%
  \BibitemOpen
  \bibfield  {author} {\bibinfo {author} {\bibfnamefont {J.~P.}\ \bibnamefont
  {Perdew}}, \bibinfo {author} {\bibfnamefont {A.}~\bibnamefont {Ruzsinszky}},
  \bibinfo {author} {\bibfnamefont {G.~I.}\ \bibnamefont {Csonka}}, \bibinfo
  {author} {\bibfnamefont {L.~A.}\ \bibnamefont {Constantin}}, \ and\ \bibinfo
  {author} {\bibfnamefont {J.}~\bibnamefont {Sun}},\ }\href@noop {} {\bibfield
  {journal} {\bibinfo  {journal} {Phys. Rev. Lett.}\ }\textbf {\bibinfo
  {volume} {103}},\ \bibinfo {pages} {026403} (\bibinfo {year}
  {2009})}\BibitemShut {NoStop}%
\bibitem [{\citenamefont {Tannor}\ \emph {et~al.}(1994)\citenamefont {Tannor},
  \citenamefont {Marten}, \citenamefont {Murphy}, \citenamefont {Friesner},
  \citenamefont {Sitkoff}, \citenamefont {Nicholls}, \citenamefont {Ringnalda},
  \citenamefont {Goddard},\ and\ \citenamefont
  {Honig}}]{solvation-exp-compiled-1}%
  \BibitemOpen
  \bibfield  {author} {\bibinfo {author} {\bibfnamefont {D.~J.}\ \bibnamefont
  {Tannor}}, \bibinfo {author} {\bibfnamefont {B.}~\bibnamefont {Marten}},
  \bibinfo {author} {\bibfnamefont {R.}~\bibnamefont {Murphy}}, \bibinfo
  {author} {\bibfnamefont {R.~A.}\ \bibnamefont {Friesner}}, \bibinfo {author}
  {\bibfnamefont {D.}~\bibnamefont {Sitkoff}}, \bibinfo {author} {\bibfnamefont
  {A.}~\bibnamefont {Nicholls}}, \bibinfo {author} {\bibfnamefont
  {M.}~\bibnamefont {Ringnalda}}, \bibinfo {author} {\bibfnamefont {W.~A.}\
  \bibnamefont {Goddard}}, \ and\ \bibinfo {author} {\bibfnamefont
  {B.}~\bibnamefont {Honig}},\ }\href@noop {} {\bibfield  {journal} {\bibinfo
  {journal} {J. Am. Chem. Soc.}\ }\textbf {\bibinfo {volume} {116}},\ \bibinfo
  {pages} {11875} (\bibinfo {year} {1994})}\BibitemShut {NoStop}%
\bibitem [{\citenamefont {Marten}\ \emph {et~al.}(1996)\citenamefont {Marten},
  \citenamefont {Kim}, \citenamefont {Cortis}, \citenamefont {Friesner},
  \citenamefont {Murphy}, \citenamefont {Ringnalda}, \citenamefont {Sitkoff},\
  and\ \citenamefont {Honig}}]{solvation-exp-compiled-2}%
  \BibitemOpen
  \bibfield  {author} {\bibinfo {author} {\bibfnamefont {B.}~\bibnamefont
  {Marten}}, \bibinfo {author} {\bibfnamefont {K.}~\bibnamefont {Kim}},
  \bibinfo {author} {\bibfnamefont {C.}~\bibnamefont {Cortis}}, \bibinfo
  {author} {\bibfnamefont {R.~A.}\ \bibnamefont {Friesner}}, \bibinfo {author}
  {\bibfnamefont {R.~B.}\ \bibnamefont {Murphy}}, \bibinfo {author}
  {\bibfnamefont {M.~N.}\ \bibnamefont {Ringnalda}}, \bibinfo {author}
  {\bibfnamefont {D.}~\bibnamefont {Sitkoff}}, \ and\ \bibinfo {author}
  {\bibfnamefont {B.}~\bibnamefont {Honig}},\ }\href@noop {} {\bibfield
  {journal} {\bibinfo  {journal} {J. Phys. Chem.}\ }\textbf {\bibinfo {volume}
  {100}},\ \bibinfo {pages} {11775} (\bibinfo {year} {1996})}\BibitemShut
  {NoStop}%
\bibitem [{\citenamefont {Kelly}\ \emph {et~al.}(2006)\citenamefont {Kelly},
  \citenamefont {Cramer},\ and\ \citenamefont {Truhlar}}]{solvation-exp-ions}%
  \BibitemOpen
  \bibfield  {author} {\bibinfo {author} {\bibfnamefont {C.~P.}\ \bibnamefont
  {Kelly}}, \bibinfo {author} {\bibfnamefont {C.~J.}\ \bibnamefont {Cramer}}, \
  and\ \bibinfo {author} {\bibfnamefont {D.~G.}\ \bibnamefont {Truhlar}},\
  }\href@noop {} {\bibfield  {journal} {\bibinfo  {journal} {J. Phys. Chem. B}\
  }\textbf {\bibinfo {volume} {110}},\ \bibinfo {pages} {16066} (\bibinfo
  {year} {2006})}\BibitemShut {NoStop}%
\bibitem [{\citenamefont {Needs}\ \emph {et~al.}(2010)\citenamefont {Needs},
  \citenamefont {Towler}, \citenamefont {Drummond},\ and\ \citenamefont
  {R\'ios}}]{CASINO}%
  \BibitemOpen
  \bibfield  {author} {\bibinfo {author} {\bibfnamefont {R.~J.}\ \bibnamefont
  {Needs}}, \bibinfo {author} {\bibfnamefont {M.~D.}\ \bibnamefont {Towler}},
  \bibinfo {author} {\bibfnamefont {N.~D.}\ \bibnamefont {Drummond}}, \ and\
  \bibinfo {author} {\bibfnamefont {P.~L.}\ \bibnamefont {R\'ios}},\
  }\href@noop {} {\bibfield  {journal} {\bibinfo  {journal} {J. Phys.: Condens.
  Matter}\ }\textbf {\bibinfo {volume} {22}},\ \bibinfo {pages} {023201}
  (\bibinfo {year} {2010})}\BibitemShut {NoStop}%
\bibitem [{\citenamefont {Trail}\ and\ \citenamefont
  {Needs}(2005{\natexlab{a}})}]{TrailNeeds}%
  \BibitemOpen
  \bibfield  {author} {\bibinfo {author} {\bibfnamefont {J.}~\bibnamefont
  {Trail}}\ and\ \bibinfo {author} {\bibfnamefont {R.}~\bibnamefont {Needs}},\
  }\href@noop {} {\bibfield  {journal} {\bibinfo  {journal} {J. Chem. Phys.}\
  }\textbf {\bibinfo {volume} {122}},\ \bibinfo {pages} {174109} (\bibinfo
  {year} {2005}{\natexlab{a}})}\BibitemShut {NoStop}%
\bibitem [{\citenamefont {Trail}\ and\ \citenamefont
  {Needs}(2005{\natexlab{b}})}]{TN2}%
  \BibitemOpen
  \bibfield  {author} {\bibinfo {author} {\bibfnamefont {J.}~\bibnamefont
  {Trail}}\ and\ \bibinfo {author} {\bibfnamefont {R.}~\bibnamefont {Needs}},\
  }\href@noop {} {\bibfield  {journal} {\bibinfo  {journal} {J. Chem. Phys.}\
  }\textbf {\bibinfo {volume} {122}},\ \bibinfo {pages} {014112} (\bibinfo
  {year} {2005}{\natexlab{b}})}\BibitemShut {NoStop}%
\end{thebibliography}
%

\end{document}